\documentclass[12pt]{iopart}
\usepackage{iopams}
\expandafter\let\csname equation*\endcsname\relax
\expandafter\let\csname endequation*\endcsname\relax
\usepackage{amsmath}
\usepackage{amssymb}
\usepackage{graphicx}
\usepackage{epstopdf}
\begin{document}
\title {Quantum thermodynamic properties of a cold atom coupled to a heat bath in non-Abelian gauge potentials}
\vskip 0.5cm \author{Asam Rajesh$^1$, Malay Bandyopadhyay$^{1,\dagger}$ and A. M. Jayannavar$^{2,3}$}
\vskip 0.5cm
\address{1. School of Basic Sciences, Indian Institue of Technology Bhubaneswar, Bhubaneswar, India 751007\\
2. Institute of Physics, Sachivalaya Marg, Sainik School PO, Bhubaneswar, India, 751005\\
3.Homi Bhabha National Institute, Training School Complex, Anushakti Nagar,
Mumbai-400085, India}
\ead{$\dagger$ malay@iitbbs.ac.in}
\vskip 0.5cm
\begin{abstract}
In this work, we study different quantum thermodynamic functions (QTFs) of a cold atom subjected to an artificial non-Abelian uniform magnetic field and linearly coupled to a quantum heat bath through either usual coordinate-coordinate coupling or through momentum variables. The bath is modelled as a collection of independent quantum harmonic oscillators. In each of the coupling scheme, the effect of the non-Abelian magnetic field on different QTFs are explicitly demonstrated for a U(2) gauge transformation. In each case, we show that the free energy has a different expression than that for the Abelian case. We consider two illustrative heat bath spectrum (Ohmic bath and Drude model) to evaluate explicit closed form expressions of free energy (F), specific heat (C), and entropy (S) in the low temperature limit for each of the above mentioned coupling scheme. The dependence of different QTFs on the non-Abelian magnetic field are pointed out even if the gauge potential is uniform in space.
\vskip 0.5cm
\end{abstract}
\pacs{05.30.-d, 05.10.Gg, 05.40.-a,05.60.Gg}
\maketitle
\section{Introduction}
The research on the dynamics of complex many-body quantum systems is becoming realizable nowadays by utilizing the method of quantum Simulation with ultra cold atoms \cite{a}. This quantum simulator suggests certain model Hamiltonians which are originally introduced to explain rich quantum behaviour in condensed matter systems \cite{b,c} and sometimes in high energy physics \cite{d}. Such model Hamiltonians are now realizable with ultra cold atoms in various type of optical potentials \cite{e}. These ultra cold atomic systems are highly controllable and are able to mimic much more complex quantum systems. One such example is quantum gas which has high flexibility, controllability, and scalability \cite{b,c}. Since, these charge neutral systems are unaffected by external electromagnetic fields, the simulation of electromagnetic condensed matter phenomena (e.g. spin Hall effect) is not realizable. So, one may suggest artificial creation of elactromagnetic fields for neutral atoms. Several suggestions can be found in literature to generate such artificial electromagnetic potentials which can exactly mimic the complex dynamics of charged particles in real electromagnetic field \cite{d}. \\
\indent
In nature Electromagnetic fields are usually known as Abelian gauge field. On the other hand, non Abelian gauge fields are responsible for the strong and weak interactions. Thus, gauge fields play crucial role  very much to understand three fundamental interactions. Recent success in simulation of Abelian and non Abelian gauge fields for ultra cold neutral atoms \cite{e,f,g,h} open the doorway of a new field for cold atom physics. In addition the detrimental effect of quantum dissipation due to its interaction with environment can be of great importance. Recently, Klein and Jaksch proposed a new mechanism to create an artificial magnetic field in an optical lattice by utilizing a rotating Bose-Einstein condensate (BEC) \cite{i}. As the temperature of a BEC is about $\mu K$, one can expect frictional force as well as random force. Thus, the effect of quantum dissipation in such system can be of great interest.\\
\indent
Using quantum Langevin method, the quantum dynamics of a cold atom in a non-Abelian magnetic field and linearly coupled to a heat bath is studied by Guingarey and Avossevou \cite{guin1}. They have pointed out the significant effect of non-Abelian magnetic field on the localization of the cold atom as well as on the average kinetic energy of the dissipative cold atom. For this purpose, they considered two relevant coupling models : the independent oscillator (IO) or coordinate-coordinate coupling scheme \cite{ford1} and the momentum-momentum coupling scheme \cite{anker, malay,cuccoli}. Further, they have derived an exact formula for the free energy of the cold atom in a non-Abelian magnetic field and in thermal equilibrium with an heat bath which is connected with the heat bath through coordinate-coordinate scheme \cite{guin2}. In this present analysis, we go further by deriving an exact formula for the free energy of the same system with momentum-momentum coupling. Not only that, we derive closed form expressions of free-energy, entropy and specific heat for two different heat bath spectrum of (Ohmic and Drude) at low temperatures and explicitly demonstrate the effect of non-Abelian magnetic field on these thermodynamic quantities for the both coupling schemes : coordinate-coordinate  and momentum-momentum schemes. In each case, we derive the exact formula for the free energy for $U(n)$ gauge transformation. Then, the simplest $U(2)$ gauge transformation is taken into account to demonstrate the explicit contribution of non-Abelian magnetic field in different thermodynamic quantities.\\
\indent
The paper is organized as follows. In the following section, we describe our model of study for the coordinate-coordinate coupling schemes. In section 2, we also derive closed form expressions of free energy (F), entropy (S) and specific heat (C) at low temperatures for Ohmic heat bath spectrum as well as for the Drude model in the coordinate-coordinate coupling scheme. In section 3, we establish an exact formula for the equilibrium free energy of the cold atom in the momentum-momentum coupling model. Then we follow the same analysis of different thermodynamic quantities as mentioned for section 2. In section 4, we analyze the effect of non-Abelian dynamics on the magnetic moment originating from the internal degrees of freedom of the cold atom . We conclude our paper in section 5.
\section{coordinate-coordinate coupling}
We consider a cold atom in a harmonic potential in the presence of an artificial non-Abelian magnetic field. The cold atom is linearly coupled through the coordinate variables to a large number $\mathcal{N}$ of independent quantum harmonic oscillators constituting a heat bath. The Hamiltonian of the system is given by
\begin{eqnarray}
 H_{c-c}&=&\frac{1}{2m}\Big( p_{\alpha}\mathcal{I}_n-\frac{e}{c}A_{\alpha}\Big)^2+\frac{1}{2}m\omega_0^2r_\alpha^2 \mathcal{I}_n \nonumber \\ &+&\sum_{j=1}^{\mathcal{N}}\Big\lbrack\frac{1}{2m_j}p_{j\alpha}^2\mathcal{I}_n+\frac{1}{2}m_j\omega_j^2(q_{j\alpha}\mathcal{I}_n-r_\alpha\mathcal{I}_n)^2\Big\rbrack
\end{eqnarray}	
In this expression of Hamiltonian, m, $r_\alpha\mathcal{I}_n$ and $p_\alpha\mathcal{I}_n$ are the mass and the
components of position and the momentum operators of the cold atom, respectively. On the other hand, $\omega_0$ is the frequency characterizing its motion in the harmonic well. The $j$th heat-bath oscillator is characterized by its mass $m_j$, and frequency $\omega_j$, while $q_{j\alpha}\mathcal{I}_n$ and $p_{j\alpha}\mathcal{I}_n$ are, respectively, components of the position and the conjugate momentum operators for the jth heat-bath oscillator. The velocity of the light in the vacuum is denoted by c. In the context of artificial non-Abelian magnetic field, the parameter e is referred as artificial charge as mentioned in Ref. \cite{j}. Here, $\mathcal{I}_n$ is a $n\times n$ identity matrix in the form of a $U(n)$ group which will be specified later. Throughout this paper, we denote three spatial directions by Greek indices $(\alpha,\beta,\gamma...)$ and Roman indices $(i,j,k....)$ represent the heat bath oscillators. We follow Einstein summation convention for the Greek indices.\\
\indent
The components of vector gauge potential $A_\alpha$ which are denoted by $n\times n$ Hermitian matrix and are related to the corresponding  magnetic field $\mathcal{B}$ as follows,
\begin{equation}
\vec{\mathcal{B}}=\vec{\triangledown}\times\vec{A}-\frac{ie}{\hbar c}\vec{A}\times\vec{A}
\end{equation}
The first term in the right-hand side of Eq. (2) is the usual Abelian contribution, while the second term takes care of the non-Abelian contribution to the magnetic field and its components are given by
\begin{equation}
\mathcal{B}_\alpha=\epsilon_{\alpha\beta\gamma}\Big(\partial_\beta A_\gamma-\frac{ie}{\hbar c}A_\beta A_\gamma\Big),
\end{equation}
where,	$\epsilon_{\alpha\beta\gamma}$ denotes Levi-Civita tensor. The second term identically vanishes for the Abelian case. Unlike Abelian case, a uniform gauge potential $\vec{A}$ can create a non-zero magnetic field. It is evident that one can decompose the magnetic field $\vec{\mathcal{B}}$ in Eq. (3) into two components for a $U(n)$ gauge group as follows :
\begin{equation}
\vec{\mathcal{B}}=\vec{B}+\vec{N},
\end{equation}
where, $\vec{B}=(B_x \mathcal{I}_n,B_y \mathcal{I}_n,B_z \mathcal{I}_n)$ is the Abelian part and $\vec{N}=(N_x,N_y,N_z)$ is the non-Abelian contribution. The components of $\vec{N}$ can be represented as follows
\begin{equation}
N_{\rho}=\Lambda_{\rho,\beta}\sigma_{\beta} \mathcal{I}_n.
\end{equation}
Here, $\Lambda_{\rho,\beta}$ and $\sigma_{\beta}$ are real numbers and Pauli matrices, respectively and the magnitude of the non-Abelian part of the magnetic field is $N = \sqrt{N_x^2 + N_y^2+N_z^2}$, with $N_{\rho}^2=(\Lambda_{\rho,x}^2+\Lambda_{\rho,x}^2+\Lambda_{\rho,x}^2)\mathcal{I}_n$.
Various relevant commutation relations for the different coordinate and momentum operators are
\begin{equation}
[r_\alpha,p_\beta]=i\hbar \delta_{\alpha\beta}\mathcal{I}_n, [q_{j\alpha},p_{k\beta}]=i\hbar \delta_{jk} \delta_{\alpha\beta}\mathcal{I}_n
\end{equation}
while all other commutators vanish. In the above equation, $\delta_{jk}$ denotes the Kronecker delta function. Now, we introduce the covariant derivatives as
\begin{equation}
D_\alpha=\partial_\alpha-\frac{ie}{\hbar c}A_\alpha=\frac{i}{\hbar}\Big(p_\alpha-\frac{e}{c}A_\alpha\Big),
\end{equation}
which can help to express the component of the magnetic field as follows:
\begin{equation}
\mathcal{B}_\alpha=\epsilon_{\alpha\beta\gamma}D_\beta A_\gamma
\end{equation}
With the help of the covariant derivative expression (Eq.7), we can rewrite the Hamiltonian Eq.(1) as follows :
\begin{equation}
H_{c-c}=\frac{-\hbar^2}{2m}D_\alpha^2+\frac{1}{2}m\omega_0^2r_\alpha^2\mathcal{I}_n +\sum_{j=1}^{\mathcal{N}}\Big\lbrack\frac{1}{2m_j}p_{j\alpha}^2\mathcal{I}_n+\frac{1}{2}m_j\omega_j^2(q_{j\alpha}\mathcal{I}_n-r_\alpha\mathcal{I}_n)^2\Big\rbrack	\end{equation}
Now, under a gauge transformation, which is nothing but the introduction of a unitary transformation U and a transformation of $A_{\alpha}$ such that
\begin{equation}
A_{\alpha}= UA_{\alpha}U^{\dagger}+\frac{i\hbar c}{e}U\partial_{\alpha}U^{\dagger},
\end{equation}
 and it results in $D_{\alpha}$ and $\mathcal{B}_{\alpha}$ transformed covariantly. Hence the Hamiltonian (Eq. 9) becomes a covariant quantity.
\subsection{Different QTF : Ohmic bath}
In Ref.\cite{guin1}, it is already shown that the free energy of a cold atom in contact with a heat bath with coordinate-cordinate coupling and in the presence of a U(2) non-Abelian gauge field is given by :
\begin{equation}
F_0(T,N)=F_0(T,0)+ \Delta F_0(T,N),
\end{equation}	
where, $F_0(T,0)$ is the free energy of the cold-atom in the absence of the magnetic field and is given by :
\begin{equation}
F_0(T,0)=\frac{3}{\pi}\int_{0}^{\infty}d\omega f(\omega,T)\Im \lbrack \frac{d}{d\omega}\ln \alpha^{(0)}(\omega)\rbrack \mathcal{I}_2,
\end{equation}
where, $\alpha^{(0)}(\omega)=1/\lambda(\omega)=m(\omega_0^2-\omega^2)-i\omega \bar{u}(\omega)$ is the susceptibility in the absence of the magnetic field. On the other hand, the second term $\Delta F_0(T,N)$ denotes the magnetic field contribution (including the non-Abelian gauge field) to the free energy of the cold atom and is given by
\begin{equation}
\Delta F_0(T,N)= -\frac{1}{2\pi}\int_{-\infty}^{+\infty}d\omega f(\omega,T) \Im \lbrace \frac{d}{d\omega}\ln \lbrack 1- \Big(\frac{\omega e\vec{N}}{c}\Big)^2\lbrack \alpha^{(0)}\rbrack^2+2\Lambda \Big(\frac{\omega e}{c}\Big)^3\lbrack \alpha^{(0)}\rbrack^3 \Big\rbrack \Big\rbrace \mathcal{I}_2,
\end{equation}
with the free energy of a free oscillator of frequency $\omega$,
\begin{eqnarray}
f(\omega,T)&=&k_BT \ln \Big\lbrack 2sinh \Big (\frac{\hbar \omega}{2k_BT}\Big)\Big\rbrack \nonumber \\
&=& \frac{\hbar \omega}{2}+k_BT\ln \Big\lbrack 1- \exp\Big(-\frac{\hbar\omega}{k_BT}\Big)\Big\rbrack
\end{eqnarray}
Then, they analyze the influence of the non-Abelian magnetic field on the magnetic moment of the cold atom. They consider a non-Abelian magnetic field obtained from the laser methods which generates degenerate dark states \cite{dark}. This tripod scheme creates a pair of such degenerate dark states which may lead to a SU(2) gauge potential if it generates a uniform vector potential in space. Following Ref. \cite{dark}, one may consider $|1>$, $|2>$, and $|3>$  as three atomic ground states which are coupled to one single excited state $|0>$. Now, for a particular orientation of laser field in which the level $1>$ and level $|2>$ are copropagating and have the same frequency and same angular momentum, the $\Lambda_{\rho,x}=0=\Lambda_{\rho,z}$. as a result of that the cubic term in the free energy expression $\Delta F_0(T,N)$ vanishes and it is reduced to
\begin{equation}
\Delta F_0(T,N)= -\frac{1}{2\pi}\int_{-\infty}^{+\infty}d\omega f(\omega,T) \Im \Big\lbrace \frac{d}{d\omega}\ln \Big\lbrack 1- \Big(\frac{\omega e\vec{N}}{c}\Big)^2\lbrack \alpha^{(0)}\rbrack^2 \Big\rbrack \Big\rbrace \mathcal{I}_2,
\end{equation}
Now, we consider a special case of the Ohmic heat bath for which we have $\bar{u}(\omega)=m\gamma$. For this special heat bath spectrum, we have the free energy
 \begin{eqnarray}
 \hskip-1.0in
 &&F_0(T,\vec{N})=\frac{3k_BT}{\pi}\int_0^{\infty}d\omega\ln\Big(1-e^{-{\hbar\omega}/{k_BT}}\Big)\Big(\frac{\omega_1}{\omega^2+\omega_1^2}+\frac{\omega_1^*}{\omega^2+{\omega_1^*}^2}\Big)\nonumber \\
 \hskip-1.0in
 &&+\frac{k_BT}{\pi}\int_0^{\infty}d\omega\ln\Big(1-e^{-{\hbar\omega}/{k_BT}}\Big)\Big(
 \frac{\Omega_1}{\omega^2+\Omega_1^2}+\frac{\Omega_1^*}{\omega^2+{\Omega_1^*}^2}+\frac{\Omega_2}{\omega^2+{\Omega_2}^2}+\frac{\Omega_2^*}{\omega^2+{\Omega_2^*}^2}\Big)\nonumber \\
 \end{eqnarray}
 where
 \begin{eqnarray}
 &&\omega_1=\frac{\gamma}{2}+i\sqrt{\omega_0^2-\frac{\gamma^2}{4}}\\\nonumber
&& \Omega_1=\Big\lbrack\frac{\gamma}{2}+\Big(\frac{b-a}{2}\Big)^{\frac{1}{2}}\Big\rbrack-i\Big\lbrack\frac{\omega_c^{na}}{2}+\Big(\frac{b+a}{2}\Big)^{\frac{1}{2}}\Big\rbrack \\ \nonumber
  && \Omega_2=\Big\lbrack\frac{\gamma}{2}-\Big(\frac{b-a}{2}\Big)^{\frac{1}{2}}\Big\rbrack-i\Big\lbrack\frac{\omega_c^{na}}{2}-\Big(\frac{b+a}{2}\Big)^{\frac{1}{2}}\Big\rbrack \\ \nonumber
  && a=\Big(\frac{\omega_c^{na}}{2}\Big)^2+\Big(\omega_0^2-\frac{\gamma^2}{4}\Big)\\ \nonumber
 && b=\Big\lbrack a^2+\Big(\frac{\gamma\omega_c^{na}}{2}\Big)^2\Big\rbrack^{\frac{1}{2}}
 \end{eqnarray}
 $\omega_1^*$,$\Omega_1^*$, and $\omega_2^*$ are the complex conjugates of $\omega_1$,$\Omega_1$, and $\omega_2$ respectively and $\omega_c^{na}=\frac{e}{mc}\frac{2e}{\hbar c}N$.
One can rewrite Eq. (16) in terms of Stieltjes function ($J(z)$) as follows :
\begin{eqnarray}
\hskip-1.0in
F(T,\vec{N})&=&3k_BT\Big\lbrack J\Big(\frac{\hbar\omega_1}{2\pi k_BT}\Big)+J\Big(\frac{\hbar\omega_1^*}{2\pi k_BT}\Big)
\Big\rbrack-k_BT\Big\lbrack
J\Big(\frac{\hbar\Omega_1}{2\pi k_BT}\Big)+J\Big(\frac{\hbar\Omega_1^*}{2\pi k_BT}\Big)\nonumber \\
\hskip-1.0in
&&+J\Big(\frac{\hbar\Omega_2}{2\pi k_BT}\Big)+J\Big(\frac{\hbar\Omega_2^*}{2\pi k_BT}\Big)\Big\rbrack,
\end{eqnarray}
where Stieltjes $J$ function is given by
\begin{equation}
J(z)=-\frac{1}{\pi}\int_0^{\infty}dt\ln\Big(1-e^{-2\pi t}\Big)\frac{z}{t^2+z^2}.
\end{equation}
\vskip 0.5cm
\subsubsection{Low-temperature expansion ($k_BT<<\hbar\omega_0$)}
In the low-temperature case, we use the asymptotic expansion for $J(z)$ :
\begin{equation}
J(z)=\sum_{n=0}^{\infty}\frac{D_{2n+2}}{(2n+1)(2n+2)}\frac{1}{z^{2n+1}},
\end{equation}
with $D_2=\frac{1}{6}$; $D_4=-\frac{1}{30}$; $D_6=\frac{1}{42}$; $D_8=-\frac{1}{30}$ and so on.
With this asymptotic expansion, the free energy becomes :
\begin{eqnarray}
\hskip-1.0cm
F_0(T,\vec{N})&=&-\Big\lbrack\frac{\pi(k_BT)^2\gamma}{2\hbar\omega_0^2}+\frac{\pi^3(k_BT)^4}{45\hbar^3\omega_0^6}(A_1+A_2+A_3)\nonumber \\
&&+\frac{8\pi^5(k_BT)^6(B_1+B_2+B_3)}{315\hbar^5\omega_0^{10}}+..\Big\rbrack,
\end{eqnarray}
where $A_1=\gamma(3\omega_0^2-\gamma^2)$, $A_2=(3\Gamma_1^2\lambda_1-\Lambda_1^3)$, $A_3=(3\Gamma_2^2\lambda_2-\Lambda_2^3)$, $B_1=\gamma(5\omega_0^4-5\gamma^2\omega_0^2+\gamma^4)$, $B_2=(\lambda_1^5-10\lambda_1^3\Gamma_1^2+5\lambda_1\Gamma_1^4)$, $B_3=(\lambda_2^5-10\lambda_2^3\Gamma_2^2+5\lambda_2\Gamma_2^4)$, $\Gamma_{1,2}=\frac{\omega_c^{na}}{2}\pm\sqrt{\frac{b+a}{2}}$; $\lambda_{1,2}=\frac{\gamma}{2}\pm\sqrt{\frac{b+a}{2}}$; and $\Lambda_{1,2}=\gamma\pm \sqrt{2(b-a)}$. Entropy can be written as
\begin{eqnarray}
\hskip -0.4cm
S(T,\vec{N})&=&k_B\Big\lbrack\frac{\pi k_BT\gamma}{\hbar\omega_0^2}+\frac{4\pi^3(k_BT)^3}{45\hbar^3\omega_0^6}(A_1+A_2+A_3)\nonumber \\
\hskip-0.4cm
&&+\frac{16\pi^5(k_BT)^5(B_1+B_2+B_3)}{105\hbar^5\omega_0^{10}}+...\Big\rbrack.
\end{eqnarray}
Finally, specific heat of the system is given by
\begin{eqnarray}
\hskip-0.4cm
C(T,\vec{N})&=&k_B\Big\lbrack\frac{\pi k_BT\gamma}{\hbar\omega_0^2}+\frac{4\pi^3(k_BT)^3}{15\hbar^3\omega_0^6}(A_1+A_2+A_3)\nonumber \\
\hskip-0.4cm
&& +\frac{16\pi^5(k_BT)^5(B_1+B_2+B_3)}{21\hbar^5\omega_0^{10}}+...\Big\rbrack
\end{eqnarray}
\vskip 0.5cm
\begin{figure}
\includegraphics[width=2.0in, height=2.0in]{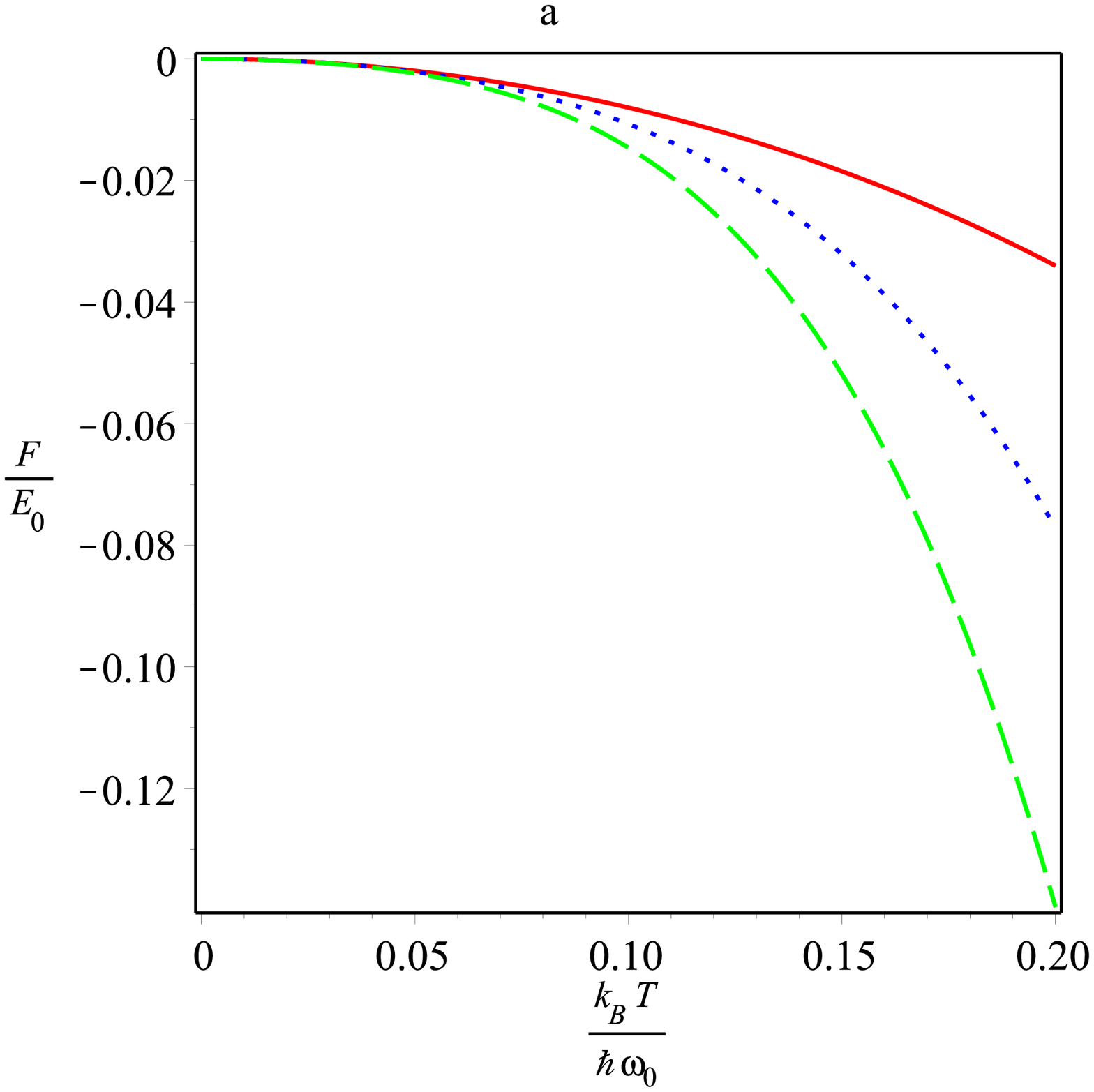}
\includegraphics[width=2.0in, height=2.0in]{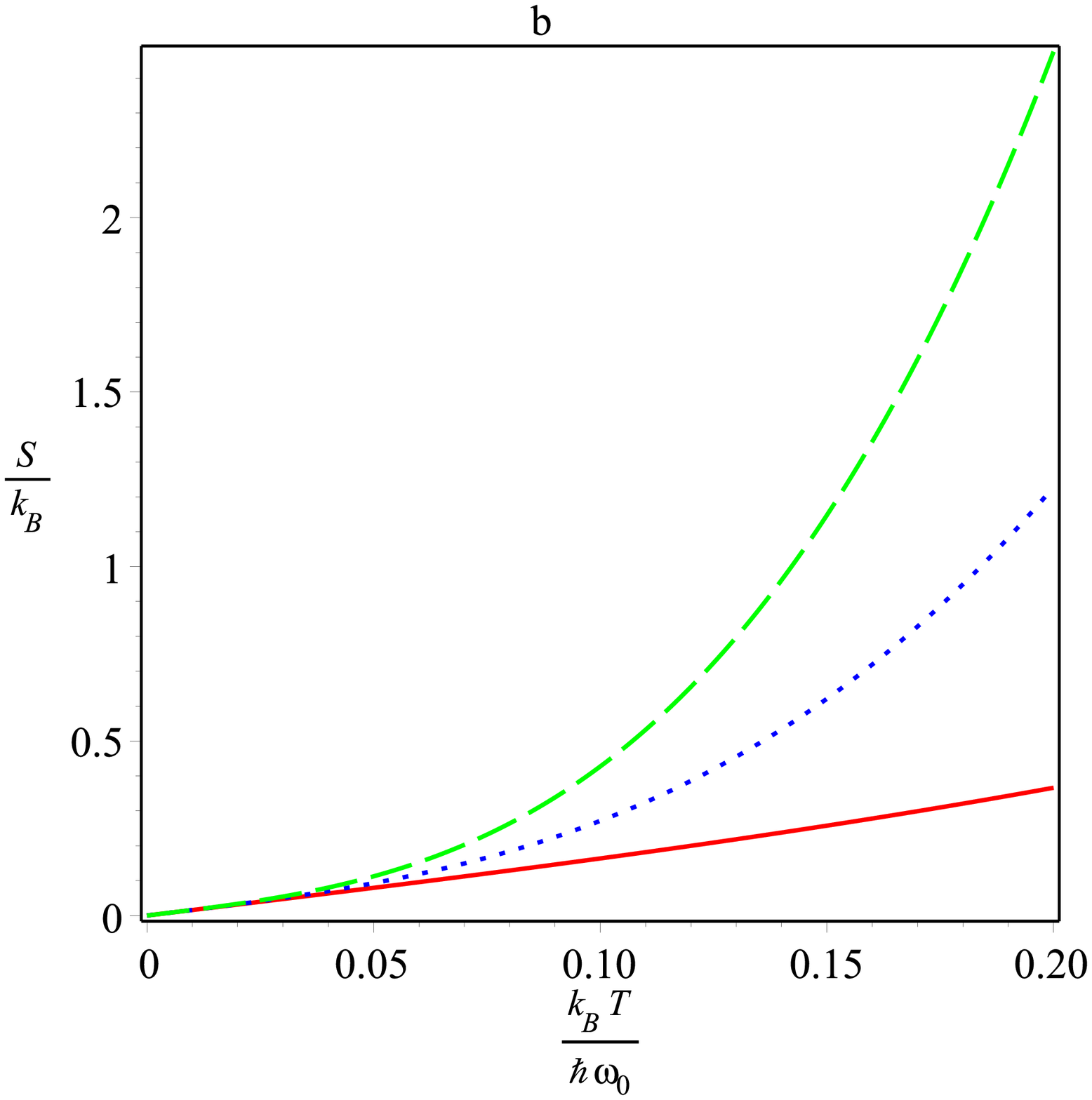}
\includegraphics[width=2.0in, height=2.0in]{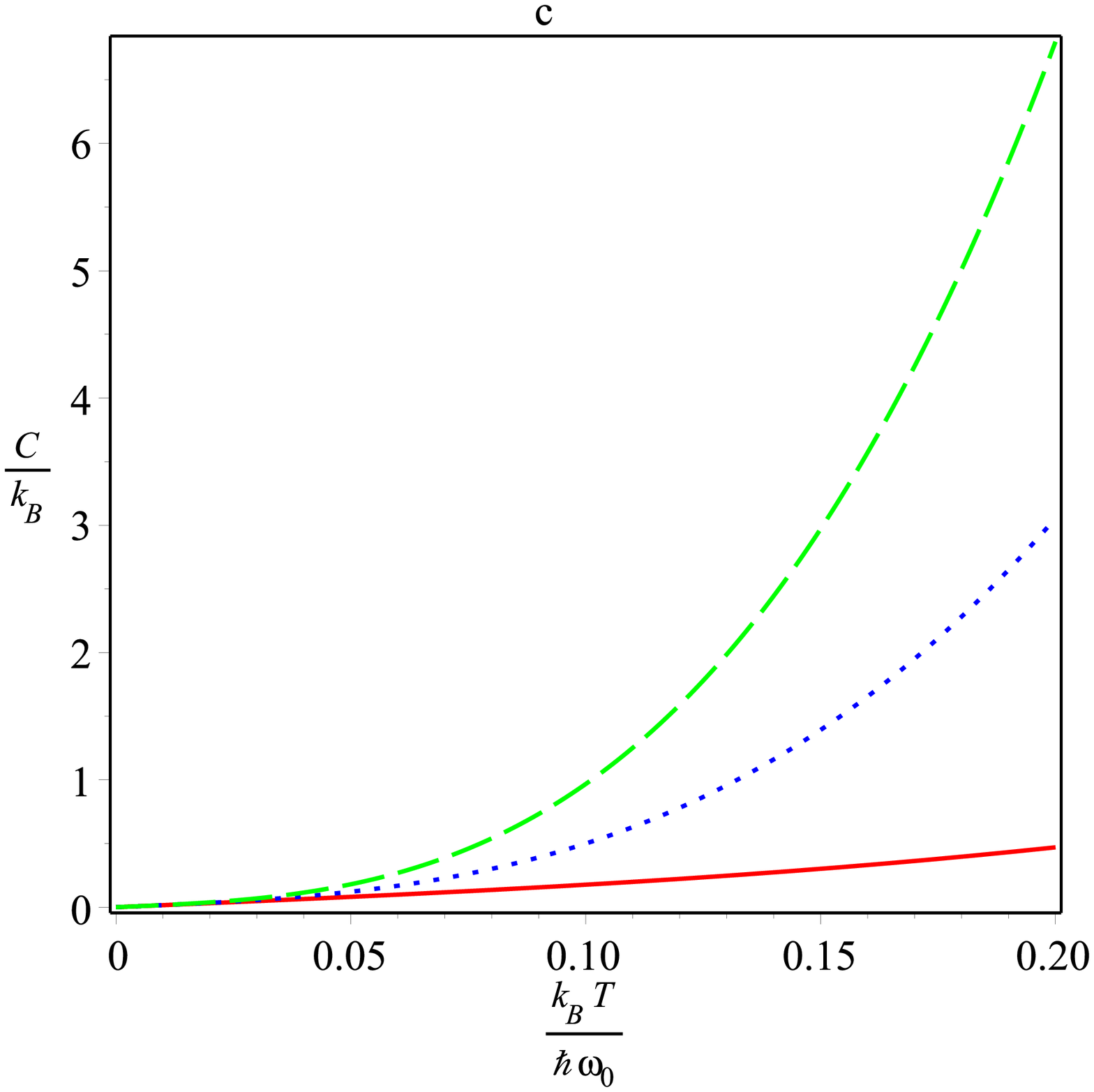}
\caption{(Color online) Plot of (a) $\frac{F}{E_{0}}$, (b)$\frac{S}{k_{B}}$, and (c)$\frac{C}{k_{B}}$, versus dimensionless temperature, $\frac{k_BT}{\hbar\omega_0}$, for the charged magneto-oscillator coupled to a Ohmic heat bath in the Low temperature regime for different values of $ N $; red solid line ($N=0$), blue dotted line ($N=1$) and green dashed line ($N=1.5$). To plot this figure, we also use $\frac{\gamma}{\omega_0}=0.5$}.
\end{figure}
To explicitly demonstrate the effect of non-Abelian magnetic field on different QTFs, we consider a particular type of uniform vector potential so that the Abelian part of the magnetic field is zero ($B_{\gamma}=0$) and the non-Abelian part of the magnetic field is along the z-axis : $\mathcal{B}=\frac{2e}{\hbar c} \Lambda \sigma_z$. Then, we derive the explicit expressions of different QTFs for the Ohmic bath in the low temperatures. We plot different quantum thermodynamic functions, like free energy $F$, entropy $S$, and specific heat $C$, as a function of dimensionless temperature $\frac{k_BT}{\hbar\omega_0}$ in Fig. 1 to demonstrate the non-Abelian effect at low temperatures. The results for different values of non-Abelian magnetic field,i.e., for different $N$ values are shown explicitly in different symbols. Our analysis clearly show the non-Abelian effect in different QTFs.\\
\subsection{Drude Model}
We consider the case with strict ohmic heat bath for our purpose. Now, we consider the case of Drude model where the memory function is regularized by introducing Drude cut-off as follows :
\begin{equation}
\tilde{u}(\nu_n)=\frac{\gamma\omega_D}{\nu_n+\omega_D}
\end{equation}
where, $\omega_D$ is the Drude cut-off frequency and it reduces to strict Ohmic case for $\omega_D\rightarrow \infty$. With this regularization in the memory function, the partition function of the dissipative cold atom system is given by \cite{jishad1,jishad2}
\begin{equation}
\mathcal{Z}=\Big(\frac{\hbar \omega_0}{4\pi^2k_BT}\Big)^2\frac{\prod_{k=1}^{3}\Gamma(\frac{\lambda_k^+}{\nu})\Gamma(\frac{\lambda_k^-}{\nu})}{\Big\lbrack\Gamma(\frac{\omega_D}{\nu})\Big\rbrack^2}
\end{equation}
where, $\Gamma(z)$ is the gamma function and its arguments can be found from the so-called Vieta equations :
\begin{eqnarray}
\lambda_1^{\pm}+\lambda_2^{\pm}+\lambda_3^{\pm}=\omega_D\pm i\omega_c^{na} \nonumber \\
\lambda_1^{\pm}\lambda_2^{\pm}+\lambda_2^{\pm}\lambda_3^{\pm}+\lambda_3^{\pm}\lambda_1^{\pm}=\omega_0^2+\gamma\omega_D\pm i\omega_D\omega_c^{na} \nonumber \\
\lambda_1^{\pm}\lambda_2^{\pm}\lambda_3^{\pm}=\omega_0^2\omega_D,
\end{eqnarray}
where, $\omega_c^{na}=\frac{e}{mc}\sqrt{\Lambda_{xy}^2+\Lambda_{yy}^2+\Lambda_{zy}^2}$.
Hence, the Helmholtz free energy of such system is given by
\begin{equation}
F=-2k_BT\ln\Big(\frac{\hbar\omega_0}{4\pi^2k_bT}\Big)-k_BT\sum_{k=1}^{3}\Big\lbrack \ln\Gamma\Big(\frac{\lambda_k^+}{\nu}\Big)+\ln\Gamma\Big(\frac{\lambda_k^-}{\nu}\Big)\Big\rbrack+2k_BT\ln\Gamma\Big(\frac{\omega_D}{\nu}\Big)
\end{equation}
The internal energy is given by
\begin{equation}
U=-2k_BT-k_BT\sum_{k=1}^{3}\Big\lbrack \frac{\lambda_k^+}{\nu}\psi\Big(\frac{\lambda_k^+}{\nu}\Big)+\frac{\lambda_k^-}{\nu}\psi\Big(\frac{\lambda_k^-}{\nu}\Big)\Big\rbrack+2k_BT\frac{\omega_D}{\nu}\psi\Big(\frac{\omega_D}{\nu}\Big),
\end{equation}
where, $\psi(z)=\frac{\partial}{\partial z}\ln \Gamma(z)$ is the digamma function.
\begin{equation}
C=2k_B+k_B\sum_{k=1}^{3}\Big\lbrack \Big(\frac{\lambda_k^+}{\nu}\Big)^2\psi^{\prime}\Big(\frac{\lambda_k^+}{\nu}\Big)+\Big(\frac{\lambda_k^-}{\nu}\Big)^2\psi^{\prime}\Big(\frac{\lambda_k^-}{\nu}\Big)\Big\rbrack-\Big(\frac{\omega_D}{\nu}\Big)^2\psi^{\prime}\Big(\frac{\omega_D}{\nu}\Big)
\end{equation}
On the other hand, the entropy is given by
\begin{equation}
S=k_B\Big\lbrace 2\lbrack \ln\Big(\frac{\hbar\omega_0}{4\pi^2k_BT}\Big)-1\rbrack +\sum_{k=1}^{3}\lbrack f\Big(\frac{\lambda_k^+}{\nu}\Big)+f\Big(\frac{\lambda_k^-}{\nu}\Big) \rbrack-2f\Big(\frac{\omega_D}{\nu}\Big)\Big\rbrace
\end{equation}
where, $f(z)=\ln \Gamma(z)-z\psi(z)$. Now, one can obtain the low temperature expressions for different QTF :
Internal energy, U :
\begin{equation}
U=\frac{\pi}{3}\hbar \gamma \Big(\frac{k_BT}{\hbar\omega_0}\Big)^2+\frac{\hbar}{2\pi}\sum_{k=1}^{3}\Big\lbrack \lambda_k^+\ln\Big(\frac{\omega_D}{\lambda_k^+}\Big)+ \lambda_k^-\ln\Big(\frac{\omega_D}{\lambda_k^-}\Big)\Big\rbrack
\end{equation}
Similarly, the low temperature expression of the specific heat given by
\begin{equation}
C=\frac{2\pi}{3}\frac{\gamma}{\omega_0^2}\frac{k_B^2T}{\hbar}+ O(T^3),
\end{equation}
and the entropy at low temperatures vanishes like
\begin{equation}
S=\frac{2\pi}{3}\frac{\gamma}{\omega_0^2}\frac{k_B^2T}{\hbar}+ O(T^3),
\end{equation}
It is to be noticed all QTFs vanishes in conformity with the third law of thermodynamics.
\section{ Momentum-momentum coupling scheme}
In this section, we take into account another type of coupling scheme for a single cold atom to a heat bath. In this complementary possibility of the coupling of a quantum system to a quantum heat bath is considered through the momentum variables. In reality, this kind of coupling scheme is employed to describe electromagnetic problems, like, quantum oscillator in black-body electromagnetic field and in superconducting quantum interference devices. In the context of a Josephson junction and an atomic Bose-Einstein condensate oscillating between two symmetric wells seperated by a barrier, the momentum-momentum coupling scheme can be applied.
\subsection{ The model and its equation of motions}
The Hamiltonian of the system is given by,
 \begin{eqnarray}
 	H_{m-m}&=&\frac{1}{2m}\Big( p_{\alpha}\mathcal{I}_n-\frac{e}{c}A_{\alpha}\Big)^2+\frac{1}{2}m\omega_0^2r_\alpha^2\mathcal{I}_n\nonumber \\ &+&\sum_{j=1}^{\mathcal{N}}\Big\lbrack\frac{1}{2m_j}\Big(p_{j\alpha}\mathcal{I}_n-g_{j}p_{\alpha}\mathcal{I}_n+\frac{g_{j}e}{c}A_{\alpha}\Big)^2+\frac{1}{2}m_j\omega_j^2q_{j\alpha}^2\mathcal{I}_n\Big\rbrack
\end{eqnarray}	
 where $g_{j}$ are the dimensionless parameters which describes the couplings, with $j=1,2,....\mathcal{N}$.
 Following Ref. \cite{guin1,malay1}, one can show that the generalized quantum Langevin equation for $U(2)$ non-Abelian gauge potential is as follows:
 \begin{eqnarray}
 \hskip-1.0in
&&m_{r}\ddot{\vec{r}}+\int_{-\infty}^{t}dt'u(t-t')\dot{\vec{r}}(t')+m_{r}\omega_0^2\vec{r}+u_d(t)\vec{r}(0)
-\frac{e}{c}(\dot{\vec{r}}\times\vec{B})+\frac{e^2}{\hbar c^2}(\dot{\vec{r}}  \Gamma_1-\Gamma_1 \dot{\vec{r}})\nonumber\\
\hskip-1.0in
 &-&\frac{e}{\hbar c}\int_{-\infty}^{t}dt'(\dot{\vec{r}}(t')  \Gamma_1-\Gamma_1 \dot{\vec{r}}(t'))u_{od}(t-t')=\vec{F}(t),
 \end{eqnarray}
 Here, the components of the magnetic field may be deccomposed in to two terms: the Abelian term, which is given by $B_\alpha$ , and the non-Abelian one, expressed as $\frac{2e}{\hbar c}\lambda_\beta^\alpha \sigma_\beta$, such that
 \begin{equation}
 \mathcal{B}_\alpha=B_\alpha+\frac{2e}{\hbar c}\lambda_\beta^\alpha \sigma_\beta
 \end{equation}
 Also, in Eq. (35), we have
 \begin{eqnarray}
 &&u(t-t')\equiv u_d(t-t')+ \Gamma_{0}u_{od}(t-t'),\\
 && F(t)=\sum_{j=1}^{\mathcal{N}} m_{j}\omega_j^2\vec{q}_{j0}(t)\vartheta(t)
 \end{eqnarray}
 where
 \begin{eqnarray}
 &&\Gamma_0\equiv
 \begin{bmatrix}
 0 & B_z & -B_y \\
 -B_z & 0 & B_x\\
 B_y & -B_x & 0
 \end{bmatrix}
 \end{eqnarray}
 \begin{eqnarray}
 &&\Gamma_1\equiv
 \begin{bmatrix}
 0 & N_z & -N_y  \\
 -N_z & 0 & N_x\\
 N_y & -N_x & 0
 \end{bmatrix}
 \end{eqnarray}
 \begin{equation}
  u_d(t-t')=\sum_{j=1}^{\mathcal{N}} \dfrac{g_{j}^2mm_{r}{\omega_{0}^2}}{m_{j}}cos[\omega_j(t-t')]\varTheta(t-t'),
 \end{equation}
 \begin{equation}
  u_{od}(t-t')=\sum_{j=1}^{\mathcal{N}} \dfrac{g_{j}^2m_{r}{\omega_{j}}e}{m_{j}c}sin[\omega_j(t-t')]\varTheta(t-t'),
 \end{equation}
 and the reduced mass $m_r$ is given by
 \begin{equation}
 m_r=\frac{m}{\Big\lbrack 1+ \sum_{j=1}^{\mathcal{N}} \frac{g_j^2m}{m_j} \Big\rbrack}
 \end{equation}
 where $\vartheta(t)$ is the Heavy side step function. In Eq.(37), the diagonal part of $u(t)$ is denoted by $u_d(t)$ and the off-diagonal part by $u_{od}(t)$. Also, one can easily notice the significant changes introduced by the non-Abelian gauge potential. Here, the additional term $\frac{e^2}{\hbar c^2}(\dot{\vec{r}}  \Gamma_1-\Gamma_1 \dot{\vec{r}})$ can be identified as the non-Abelian modification of the Lorentz force. On the other hand, the non-Abelian contribution in the frictional force is given by $\frac{e}{\hbar c}\int_{-\infty}^t dt^{\prime}(\dot{\vec{r}}(t')  \Gamma_1-\Gamma_1 \dot{\vec{r}}(t'))u_{od}(t-t')$.\\
 \subsubsection{Free energy calculation :}
 In order to compute free energy of the system, we consider a weak external c-number force $\vec{f}(t)$ to act on the relevant system particle and another set of weak external c-number forces $\vec{f}_j(t), j=1,2,....,\mathcal{N}$ to act on the bath oscillators. As a matter of fact, the perturbed Hamiltonian is given by
 \begin{equation}
 H=H_{m-m}-\vec{r}\cdot\vec{f}(t)-\sum_{j=1}^{\mathcal{N}} \vec{q}_j\cdot\vec{f}_j(t)
 \end{equation}
 In the presence of external forces $\vec{f}(t)$ and $\vec{f}_j(t)$, Eq. (35) is modified as follows :
 \begin{eqnarray}
 \hskip-1.0in
&&m_{r}\ddot{\vec{r}}+\int_{-\infty}^{t}dt'u(t-t')\dot{\vec{r}}(t')+m_{r}\omega_0^2\vec{r}+u_d(t)\vec{r}(0)\nonumber-\frac{e}{c}(\dot{\vec{r}}\times\vec{B})+\frac{e^2}{\hbar c^2}(\dot{\vec{r}}  \Gamma_1-\Gamma_1 \dot{\vec{r}})\nonumber\\
 \hskip-1.0in
 &-&\dfrac{e}{\hbar c}\int_{-\infty}^{t}dt'(\dot{\vec{r}}(t')  \Gamma_1-\Gamma_1 \dot{\vec{r}}(t'))u_{od}(t-t')=\vec{F}(t)+\vec{f}(t)+\sum_{j=1}^\mathcal{N}\int_{-\infty}^t dt^{\prime} \vec{f}_j(t^{\prime})\omega_j\sin\lbrack \omega_j(t-t^{\prime})\rbrack \Theta(t-t^{\prime}),\nonumber \\
 \hskip-1.0in
 \end{eqnarray}
 Now, we consider $SU(2)$ gauge potential, i.e. the case of a uniform vector potential in space so that the Abelian part of the magnetic field vanishes $(\vec{B}=0)$ and one can focus only on the effect of non-Abelian part of the magnetic field. In this case, our Eq. (45) reduces to
 \begin{eqnarray}
 \hskip -1.0in
 &&m_{r}\ddot{\vec{r}}+\int_{-\infty}^{t}dt'u(t-t')\dot{\vec{r}}(t')+m_{r}\omega_0^2\vec{r}+u_d(t)\vec{r}(0)+\frac{e^2}{\hbar c^2}(\dot{\vec{r}}  \Gamma_1-\Gamma_1 \dot{\vec{r}})\nonumber\\
 \hskip -1.0in
 &-&\frac{e}{\hbar c}\int_{-\infty}^{t}dt'(\dot{\vec{r}}(t')  \Gamma_1-\Gamma_1 \dot{\vec{r}}(t'))u_{od}(t-t')=\vec{F}(t)+\vec{f}(t)+\sum_{j=1}^{\mathcal{N}}\int_{-\infty}^t dt^{\prime} \vec{f}_j(t^{\prime})\eta_j(t-t^{\prime}),\nonumber \\
 \hskip -1.0in
 \end{eqnarray}
 where, $\eta_j(t-t^{\prime})=\omega_j\sin\lbrack \omega_j(t-t^{\prime})\rbrack \Theta(t-t^{\prime})$. The Fourier transform of Eq.(46) leads to the following algebraic equation,
 \begin{equation}
 \Big\lbrack\lambda(\omega) \delta_{\alpha\beta}\mathcal{I}_2+\frac{2i\omega e^2}{\hbar c^2}G({\omega})\epsilon_{\alpha\beta\gamma}N_\gamma\Big\rbrack \vec{r}_\beta=-\bar{u}r_\alpha(0)\mathcal{I}_2+\bar{F}_\alpha\mathcal{I}_2+\bar{f}_{\alpha}\mathcal{I}_2+\sum_{j=1}^{\mathcal{N}} \bar{\eta}_j\bar{f}_{j\alpha}\mathcal{I}_2,
 \end{equation}
 where,
 \begin{eqnarray}
 && \lambda(\omega)=m_{r}(\omega_0^2-\omega^2)-i\omega\bar{u}_{d}(\omega)\\
 && \bar{u}(\omega)=\int_{0}^{\infty}dt \exp(i\omega t)u(t),\\ &&\bar{r}_\beta=\int_{-\infty}^{+\infty}\exp(i\omega t)r_\beta(t).
 \end{eqnarray}
 and,
 \begin{equation}
 G({\omega})=1-\sum_{j=1}^{\mathcal{N}} \dfrac{g_{j}^2m_{r}{\omega_{j}^2}}{m_{j}(\omega_{j}^2-{\omega}^2)}
 \end{equation}
 Now, setting
 \begin{equation}
 D_{\alpha\beta}(\omega)=\lambda(\omega)\delta_{\alpha\beta}\mathcal{I}_2+\frac{2i\omega e^2}{\hbar c^2}G({\omega})\epsilon_{\alpha\beta\gamma}N_\gamma,
 \end{equation}
 One can rewrite Eq (47)
 \begin{equation}
 D_{\alpha\beta}(\omega)\bar{r}_\beta=-\bar{u}r_\alpha(0)\mathcal{I}_2+\bar{F}_\alpha\mathcal{I}_2+\vec{f}_{\alpha}\mathcal{I}_2+\sum_{j=1}^{\mathcal{N}} \bar{\eta}_j\bar{f}_{j\alpha}\mathcal{I}_2,
 \end{equation}
 From Eq.(53) we get,
 \begin{equation}
 \bar{r}_\rho=\alpha_{\rho\beta}(\omega)\lbrack -\bar{u}r_\beta(0)+\bar{F}_\beta+\bar{f}_\beta+\sum_{j=1}^{\mathcal{N}}\bar{\eta}_j\bar{f}_{j\beta}\rbrack.
 \end{equation}
 where,
 \begin{eqnarray}
 \hskip-1.0in
 &&\alpha_{\rho\beta}(\omega)\mathcal{I}_2=\lbrack D(\omega)^{-1}\rbrack_{\rho\beta} \nonumber \\
 \hskip-1.0in
 &=&\Big\lbrack\lambda^2 \delta_{\rho\beta}\mathcal{I}_2-\phi_0\Big(\frac{\omega e}{\hbar c}G(\omega)\Big)^2  N_{\rho} N_{\beta}-i\lambda\phi_0 \Big(\frac{\omega e}{ c}\Big)G(\omega)\epsilon_{\rho\beta\eta}N_{\eta}\Big\rbrack/Det D(\omega)
 \end{eqnarray}
 with, $\phi_0=\frac{e}{\hbar c}$ and
 \begin{equation}
 Det D(\omega)= \Big[\lambda^3(\omega) -\phi_0\lambda(\omega)\Big(\frac{\omega e}{c}G(\omega)\Big)^2\vec{N}^2 +2\phi_0\Big(\frac{\omega e G(\omega)}{c}\Big)^3\Lambda\Big].
 \end{equation}
  From Eq.(55), one can decompose $\alpha_{\rho\sigma}(\omega)$ in to symmetric and antisymmetric parts:
 \begin{equation}
 \alpha_{\rho\beta}(\omega)=\alpha_{\rho\beta}^s(\omega)+\alpha_{\rho\beta}^a(\omega)
 \end{equation}
 with
 \begin{equation}
 \alpha_{\rho\beta}^s(\omega)=\Big[\lambda^2 \delta_{\rho\beta}\mathcal{I}_2-\phi_0\Big(\frac{\omega e}{ c}G({\omega})\Big)^2  N_\rho N_\beta \Big] / Det D(\omega)
 \end{equation}
 and
 \begin{equation}
 \alpha_{\rho\sigma}^a(\omega)=i\lambda 2\phi_0\Big(\frac{\omega e}{c}\Big)G(\omega)\epsilon_{\rho\sigma\eta}N_{\eta} / Det D(\omega)
 \end{equation}
    Rearrnging some terms, Eq. (34) can be written in the form
    \begin{eqnarray}
     <H_{m-m}> &=& \frac{1}{2}m<\dot{r}^2>+\frac{1}{2}m\omega_0^2<{r}^2>+\Big\lbrack\sum_j g_j < \dot{r}\dot{q}_{j}+\dot{q}_{j}\dot{r}>+\sum_{j,k}g_jg_k <\dot{q}_j\dot{q}_k>\Big\rbrack\nonumber\\
    &+&\sum_j \Big(\frac{1}{2}m_j < \dot{q}_j^2>+\frac{1}{2}m_j \omega_j^2 < q_j^2>\Big)
    \end{eqnarray}
    Now, we move to the calculation of various averages. Let us first calculate the averages involving with $\vec{r}$. Following Ref. \cite{malay1}, we can write
  \begin{equation}
   \frac{1}{2}\langle \vec{r}(t)\cdot\vec{r}(t')+\vec{r}(t')\cdot\vec{r}(t)\rangle=\frac{\hbar}{\pi}\int_{0}^{\infty}d\omega G({\omega}) Im[\alpha_{\rho\rho}(\omega)]coth\Big[\frac{\hbar \omega}{2k_B T}\Big]cos[\omega(t-t')]
   \end{equation}
   Now, setting $t=t'$, we get
   \begin{equation}
    \langle {r}^2\rangle=\frac{\hbar}{\pi}\int_{0}^{\infty}d\omega G({\omega})coth\Big[\frac{\hbar \omega}{2k_B T}\Big]Im[\alpha_{\rho\rho}(\omega)].
   \end{equation}
   Differentiating Eq. (61) with respect to $t$ and $t'$ and then setting $t'$ equal to $t$, we have
   \begin{equation}
    \langle \dot{r}^2\rangle=\frac{\hbar}{\pi}\int_{0}^{\infty}d\omega G({\omega})coth\Big[\frac{\hbar \omega}{2k_B T}\Big]Im[\alpha_{\rho\rho}(\omega)]\omega^2.
   \end{equation}
   One may proceed in the similar fashion and it can be shown that similar kind of expressions hold for the averages $\langle q^2\rangle$ and $\langle \dot{q}^2\rangle$ as that of Eq. (62) and Eq. (63), but $\alpha_{\rho\rho}$ should be replaced by $\gamma_{jj,\rho\rho}$ :
   \begin{eqnarray}
   && \langle q^2\rangle=\frac{\hbar}{\pi}\int_{0}^{\infty}d\omega coth\Big[\frac{\hbar \omega}{2k_B T}\Big]Im[\gamma_{jj,\rho\rho}(\omega)]\\
   && \langle \dot{q}^2\rangle=\frac{\hbar}{\pi}\int_{0}^{\infty}d\omega coth\Big[\frac{\hbar \omega}{2k_B T}\Big]Im[\gamma_{jj,\rho\rho}(\omega)] \omega^2,
   \end{eqnarray}
   with,
   \begin{eqnarray}
   \gamma_{jk,\rho \gamma}=\frac{g_jg_kmm_r\omega^2}{m_jm_k(\omega_j^2-\omega^2)(\omega_k^2-\omega^2)}\Big(\omega_0^2\delta_{\rho\sigma}\mathcal{I}_2 +i\phi_0\frac{\omega e}{m c}\epsilon_{\rho\sigma\eta}N_{\eta} \Big)\alpha_{\sigma\gamma}(\omega)\nonumber \\
   + \frac{\delta_{jk}\delta_{\rho\gamma}}{m_k(\omega_k^2-\omega^2)}\mathcal{I}_2
   \end{eqnarray}
   and one can easily write that
   \begin{equation}
    \langle \dot{\vec{q}}_{j}\cdot\dot{\vec{q}}_{k}\rangle=\frac{\hbar}{\pi}\int_{0}^{\infty}d\omega coth\Big[\frac{\hbar \omega}{2k_B T}\Big]Im[\gamma_{jk,\rho\rho}(\omega)]\omega^2.
   \end{equation}
   Now, the averages involving both $\vec{q}_j$ and $\vec{r}$ can be written as
   \begin{equation}
      \langle \dot{\vec{q}}_{j}\cdot\dot{\vec{r}}+\dot{\vec{r}}\cdot\dot{\vec{q}}_{j}\rangle=\frac{\hbar}{\pi}\int_{0}^{\infty}d\omega coth\Big[\frac{\hbar \omega}{2k_B T}\Big]\omega^2 \Big[Im[\beta_{j,\rho\rho}(\omega)]+Im[\Delta_{j,\rho\rho}(\omega)]\Big],
   \end{equation}
  where, we have \cite{malay1}
  \begin{equation}
  \beta_{j,\rho\rho}=\frac{g_jm_r\omega^2}{m_j(\omega_j^2-\omega^2)}\alpha_{\rho\rho}(\omega),
  \end{equation}
  and
  \begin{equation}
  \Delta_{j,\rho\gamma}(\omega)=\frac{g_jmG(\omega)}{m_j(\omega_j^2-\omega^2)}\Big(\omega_0^2\delta_{\rho\sigma}+\frac{i\omega e}{mc}\epsilon_{\rho\sigma\eta}N_{\eta}\Big)\alpha_{\sigma\gamma}(\omega)-\frac{g_j}{m_j(\omega_j^2-\omega^2)}\delta_{\rho\gamma}
  \end{equation}
   We know that the mean internal energy,$U_B(T)$, of the heat bath in the absence of coupling with the cold atom is given by :
    \begin{equation}
    U_B(T)=3\sum_j \frac{\hbar \omega_j}{2} coth\Big[\frac{\hbar \omega}{2k_B T}\Big]
    \end{equation}
     On the other hand, the mean internal energy $U_0(T,\vec{N})$ of the cold atom is defined as the mean internal energy of the system of the cold atom interacting with the heat bath $\langle H_{m-m}\rangle$ minus the mean internal energy of the heat bath in the absence of coupling with the particle $U_B(T)$:
    \begin{eqnarray}
     &&U_0(T,\vec{N})=\langle H_{m-m}\rangle-U_B(T) \nonumber \\
     &=&\frac{\hbar}{2\pi}\int_{0}^{\infty}d\omega coth\Big[\frac{\hbar \omega}{2k_B T}\Big]Im \Big[\alpha_{\rho\rho}\Big(m\omega^2+m\omega_0^2+\sum_{j}\frac{m_j (\omega^2+\omega_j^2)\omega_j^2 \omega^2}{(\omega^2-\omega_j^2)^2}\Big)\Big]. \nonumber \\
    \end{eqnarray}
Therefore
 \begin{equation}
 U_0(T,\vec{N}) =\frac{1}{\pi}\int_{0}^{\infty}d\omega u(\omega,T)Im \Big[\frac{d}{d\omega}[log(Det \alpha(\omega))+\lambda(\omega)\Big(\dfrac{2i\omega e^2 \vec{N}}{\hbar c^2}\Big)^2\Big(\dfrac{d(G({\omega}))^2}{d\omega}\Big)Det \alpha(\omega)\Big]
 \end{equation}
 where $u(\omega,T)$ is the plank energy of a free oscillator of frequency $\omega$:
 \begin{equation}
 u(\omega,T)=\frac{\hbar \omega}{2}coth\Big[\frac{\hbar \omega}{2k_B T}\Big]
 \end{equation}
 The corresponding formula for the free energy of the oscillator takes the form
 \begin{equation}
 F_0(T,\vec{N}) =\frac{1}{\pi}\int_{0}^{\infty}d\omega f(\omega,T)Im \Big[\frac{d}{d\omega}[log(Det \alpha(\omega))+\lambda(\omega)\Big(\dfrac{2i\omega e^2 \vec{N}}{\hbar c^2}\Big)^2\Big(\dfrac{d(G({\omega}))^2}{d\omega}\Big)Det \alpha(\omega)\Big]
 \end{equation}
 where $f(\omega,T)$ is the free energy of a free oscillator of frequency $\omega$:
 \begin{equation}
 f(\omega,T)=kT ln[2sinh(\hbar \omega/2kT)]
 \end{equation}
 and
 \begin{equation}
 Det \alpha(\omega)=[\alpha^0(\omega)]^3 \Bigg[1-[\alpha^0(\omega)]^2\Big(\frac{2e}{\hbar c}N G({\omega})\Big)^2\Bigg]^{-1}
 \end{equation}
 so that
 \begin{equation}
 F_0(T,N)=F_0(T,0)+\Delta_1 F_0(T,N)+\Delta_2 F_0(T,N),
 \end{equation}
 where
 \begin{eqnarray}
 && F_0(T,0)=\frac{3}{\pi}\int_{0}^{\infty}d\omega f(\omega,T) Im \Big[\frac{d}{d\omega} log(Det \alpha^0 (\omega))\Big]\\
 &&\Delta_1 F_0(T,N)=-\frac{1}{\pi}\int_{0}^{\infty}d\omega f(\omega,T) Im \Big[\frac{d}{d\omega} ln\Big(1-(G({\omega})^2)[\alpha^0(\omega)]^2\Big(\frac{2e}{\hbar c}N\Big)^2 \Big(\frac{\omega e}{c}\Big)^2\Big)\Big] \nonumber \\
 \end{eqnarray}
 and
 \begin{equation}
 \Delta_2 F_0(T,N)=\frac{1}{\pi}\int_{0}^{\infty}d\omega f(\omega,T) Im \Big[\Big(\frac{2e}{\hbar c}N\Big)^2 \Big(\frac{\omega e}{c}\Big)^2\Big)\Big(\dfrac{d(G({\omega}))^2}{d\omega}\Big)\Big(1-(G({\omega})^2)[\alpha^0(\omega)]^2\Big(\frac{2e}{\hbar c}N\Big)^2 \Big(\frac{\omega e}{c}\Big)^2\Big)^{-1}\Big]
 \end{equation}
 \subsubsection{Low-temperature expansion ($k_BT<<\hbar\omega_0$)}
 We can rearrange the free energy expressions as follows :
 \begin{eqnarray}
 F_0(T,\vec{N})=F_0(T,0)+\Delta_1 F_0(T,\vec{N})+\Delta_2 F_0(T,\vec{N})\nonumber \\
 =\frac{1}{\pi}\int_0^{\infty}d\omega f(\omega,T)[3I_0-I_1+I_2],
 \end{eqnarray}
 with
 \begin{eqnarray}
 I_0&=&\Im\Big[\frac{d}{d\omega}\ln\alpha^{(0)}(\omega)\Big], \nonumber \\
 I_1&=&\Im\Big[\frac{d}{d\omega}\ln\Big\lbrace 1-(G(\omega))^2\Big(\frac{e\vec{N}\omega}{c}\Big)^2[\alpha^{(0)}(\omega)]^2\Big\rbrace\Big\rbrack, \nonumber \\
 I_2&=&\Im\Big[\frac{[\alpha^{(0)}(\omega)]^2\Big(\frac{\omega e\vec{N}}{c}\Big)^2\Big(\frac{d(G(\omega))^2}{d\omega}\Big)}{\Big\lbrace1-(G(\omega))^2\Big(\frac{e\vec{N}\omega}{c}\Big)^2[\alpha^{(0)}(\omega)]^2\Big\rbrace}\Big].
 \end{eqnarray}
 Now, $f(\omega,T)$ vanishes exponentially for $\omega>>\frac{k_BT}{\hbar}$. Therefore, in order to evaluate the free energy of the dissipative charged oscillator at low temperatures, we need to consider only low-$\omega$ contributions of integrands in evaluating the integral in Eq. (82). With this we can show that at low frequencies the magnetic field independent integrand $I_0$ becomes :
 \begin{equation}
 \lim_{\omega\rightarrow 0}I_0(\omega)\simeq \frac{C(1+\nu)}{\omega_0^2}\omega^{\nu},
 \end{equation}
 with $C=\frac{m}{m_r}b^{1-\nu}\cos\Big(\frac{\nu\pi}{2}\Big)$. On the other hand, we obtain for the  magnetic field dependent integrands $I_1$ and $I_2$ as follows  :
 \begin{eqnarray}
 \lim_{\omega\rightarrow 0}I_1= \frac{2C(\nu+1)(\omega_{cr}^{na})^2}{\omega_0^6}\omega^{\nu+2}
 =\lim_{\omega\rightarrow 0}I_2
 \end{eqnarray}
 Now, we use the result
 \begin{equation}
 \int_{0}^{\infty}dy y^{\nu}\log(1-e^{-y})=-\Gamma(\nu+1)\zeta(\nu+2),
 \end{equation}
 where $\Gamma(z)$ is the gamma function, while $\zeta(z)$ is the Riemann Zeta function, to obtain the free energy at low temperatures :
 \begin{eqnarray}
 F_0(T,0)&\simeq& -\frac{3\Gamma(\nu+2)\zeta(\nu+2)\cos\Big(\frac{\nu\pi}{2}\Big)m\hbar b}{m_r\pi}\Big(\frac{b}{\omega_0}\Big)^2\Big(\frac{k_BT}{b}\Big)^{\nu+2}\nonumber \\
 \Delta_1F_0(T,\vec{N})&\simeq& -\frac{2(\nu+1)\Gamma(\nu+3)\zeta(\nu+4)\cos\Big(\frac{\nu\pi}{2}\Big)m\hbar b}{m_r\pi}\Big(\frac{\omega_{cr}^{na}}{\omega_0}\Big)^2\Big(\frac{b}{\omega_0}\Big)^4\Big(\frac{k_BT}{b}\Big)^{\nu+4}\nonumber \\
 &=&\Delta_2F_0(T,\vec{N}).
 \end{eqnarray}
 where $\omega_{cr}^{na}=\frac{e}{m_rc}(B+\frac{2e}{\hbar c}N)$. For our uniform vector potential which is along z-direction we have $B=0$ and  $N=\sqrt{\Lambda_{xy}^2+\Lambda_{yy}^2+\Lambda_{zy}^2}$.
\begin{figure}
\includegraphics[width=2.0in, height=2.0in]{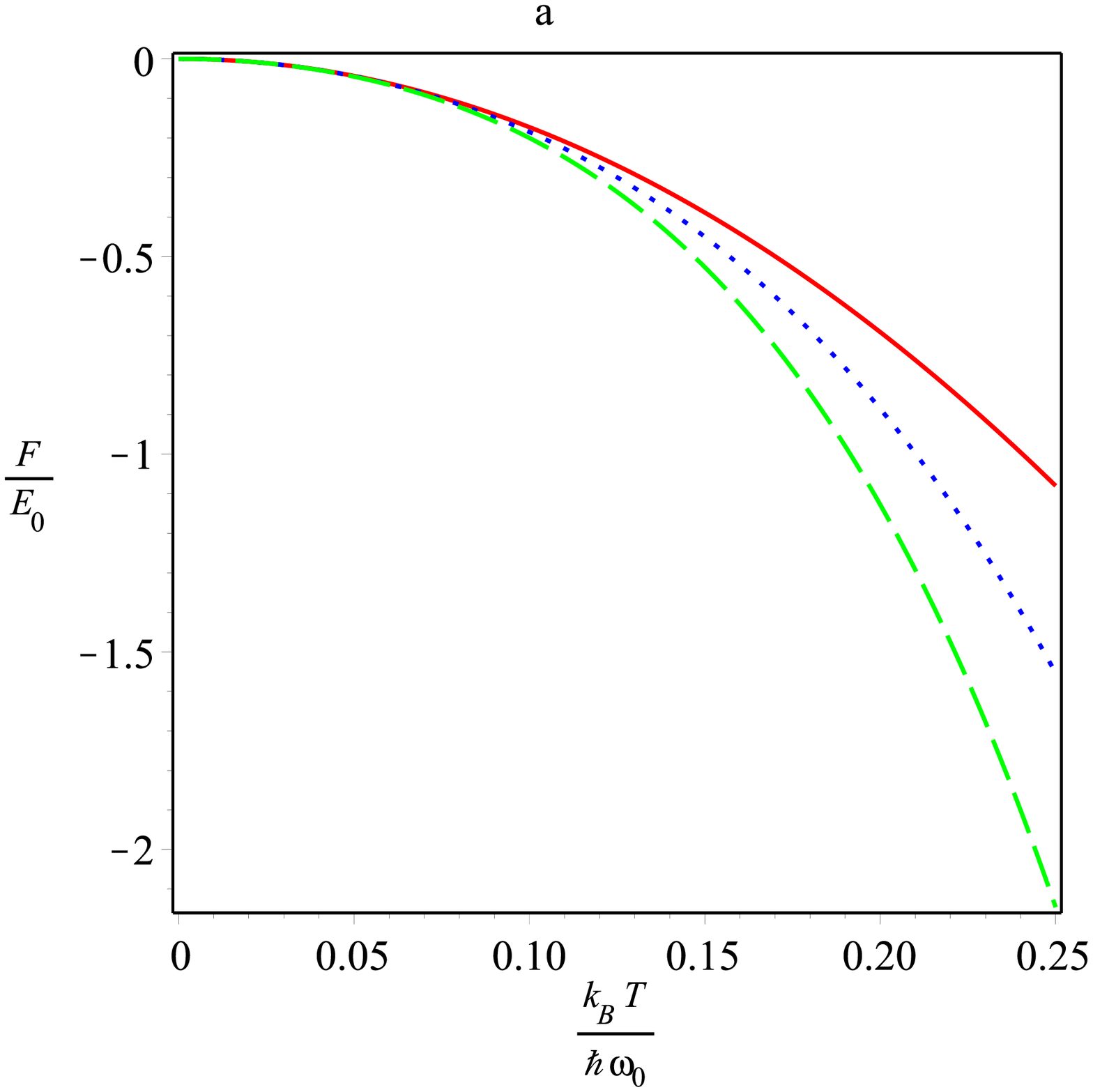}
\includegraphics[width=2.0in, height=2.0in]{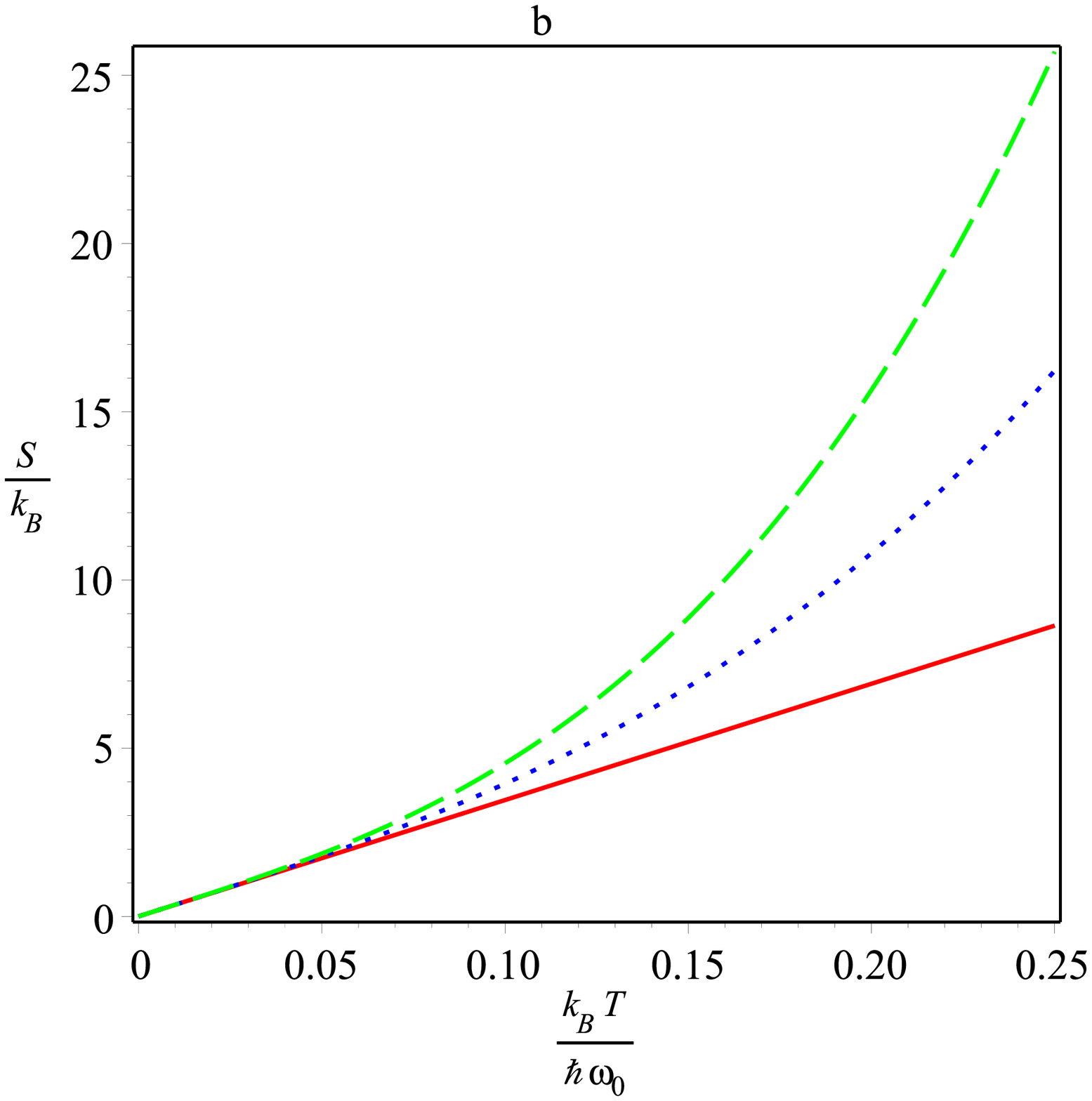}
\includegraphics[width=2.0in, height=2.0in]{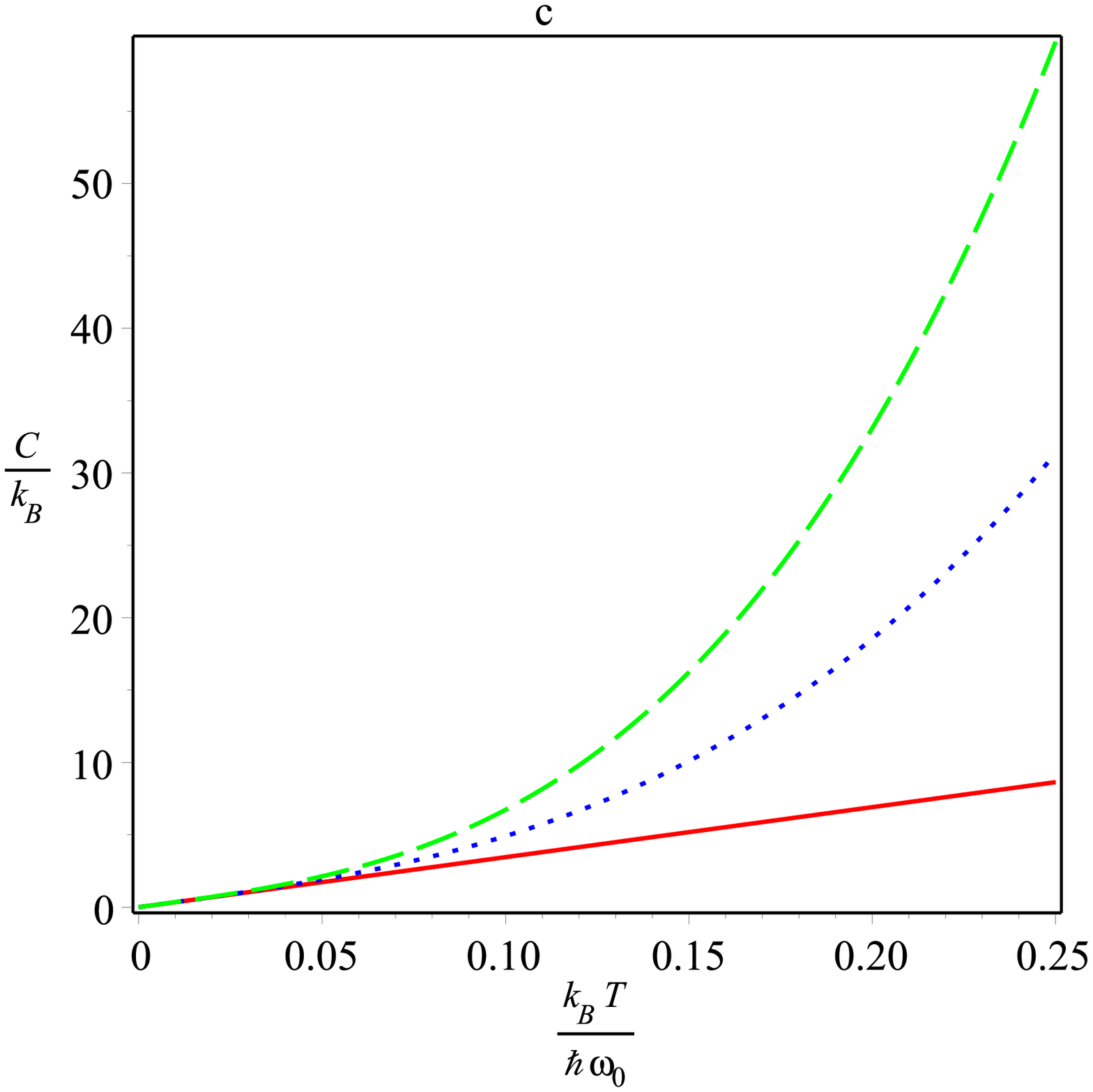}
\caption{(Color online) Plot of (a) $\frac{F}{E_{0}}$, (b)$\frac{S}{k_{B}}$, and (c)$\frac{C}{k_{B}}$, versus dimensionless temperature, $\frac{k_BT}{\hbar\omega_0}$, for the charged magneto-oscillator coupled to a Ohmic heat bath in the Low temperature regime for different values of $ N $; red solid line ($N=0$), blue dotted line ($N=1$) and green dashed line ($N=1.5$)}.
\end{figure}
\section{Influence of non-Abelian dynamics}
The main objective of this section is to analyze the effect of the non-Abelian dynamics on different QTF of the cold atom. One can basically try to find the effect of internal degrees of freedom of the cold atom on the thermodynamic functions. For this purpose, one can consider the non-Abelian magnetic field obtained from the laser methods using degenerate dark states \cite{dark}. This scheme is based on the adiabatic motion of a multilevel atom in a space dependent laser fields and it results in a space dependent dark states which gain a topological or Barry's phase. This can be thought of as the effect of an artificial magnetic field. Following Ref. \cite{guin2,dark}, we can consider the following dark states in the tripod scheme which generates SU(2) gauge with uniform vector potential :
\begin{eqnarray}
|D_1> = \sin\phi e^{iS_{31}}|1>-\cos\phi e^{iS_{32}}|2> \\
|D_2> = \cos\theta\cos\phi e^{iS_{31}}|1>+\cos\theta\sin\phi e^{iS_{32}}|2>-\sin\theta|3> ,
\end{eqnarray}
with, $S_{ij}=S_i-S_j$, $|1>$, $|2>$, and $|3>$ are three atomic ground states which are coupled to a single excited state ($|0>$) by three resonant laser fields with complex Rabi frequencies $\Omega_{\mu}$. Now. following Ref. \cite{guin2}, we can show that the free energy of the cold atom is given by
\begin{equation}
F_0(T,N)=\frac{1}{2\pi}\int_{0}^{\infty}d\omega f(\omega,T)\Im \Big\lbrack\frac{d}{d\omega}\ln\lbrack det \alpha (\omega)\rbrack \Big\rbrack\mathcal{I}_2
\end{equation}
As a matter of fact, the non-Abelian part of the magnetic moment of the cold atom in the coordinate-coordinate coupling is given by
\begin{equation}
M=-\frac{\partial F_0}{\partial N}=N\Big(\frac{e}{c}\Big)^2 \frac{\hbar}{2\pi i}\int_{-\infty}^{\infty}d\omega\omega^2\coth\Big(\frac{\hbar\omega}{2k_B T}\Big)\Big\lbrack\lambda^2 -(\frac{\omega e}{c})^2\vec{N}^2 \Big\rbrack^{-1}\mathcal{I}_2
\end{equation}
One can easily show the quantum nature of the Eq. (91), i.e. $M$ vanishes in the classical limit by expanding $\coth(\frac{\hbar \omega}{2k_BT})$ for $\hbar \rightarrow 0$ and using the analytic behaviour of the integrand in the upper half plane :
\begin{equation}
M=N\Big(\frac{e}{c}\Big)^2 \frac{k_BT}{i\pi}\int_{-\infty}^{\infty}d\omega \omega \lbrack \lambda^2 - \Big(\frac{e\omega}{c}\Big)^2\vec{N}^2\rbrack^{-1}=0 \nonumber
\end{equation}
Now, one can compute integration in Eq. (91) by choosing the contour in the upper half plane and using the following identity
\begin{equation}
\coth(z)=\sum_{n=-\infty}^{\infty}\frac{1}{z+in\pi}.
\end{equation}
 Using $\omega = i\nu_n$ and Matsubara frequencies $\nu_n=\frac{2\pi k_BTn}{\hbar}$, we can show that
\begin{equation}
M=-2k_BT\frac{e\omega_c^{na}}{mc} \sum_{n=1}^{\infty}\frac{\nu_n^2}{\hat{\lambda}^2(\nu_n)+(\nu_n\omega_c^{na})^2},
\end{equation}
with, $\hat{\lambda}(\nu_n)=\lambda(i\nu_n)/m=\omega_0^2+\nu_n^2+\nu_n\mu(i\nu_n)/m$ and the non-Abelian cyclotron frequency $\omega_c^{na}= \frac{e}{mc}\sqrt{\Lambda_{x,y}^2+\Lambda_{y,y}^2+\Lambda_{z,y}^2}$. One can easily note that this magnetic moment which originates from the motion of the centre of mass of the cold atom and its internal degrees of freedom is still negative (i.e diamagnetic) and is unaltered by the presence of the arbitrary heat bath. This fact still holds if one takes the limit $\omega_0\rightarrow 0$ in Eq. (93). Now, one can compute the explicit closed form expression of magnetic moment at zero temperature for an Ohmic heat bath :
\begin{eqnarray}
M=-\frac{\hbar}{2\pi}\Big(\frac{e}{mc}\Big)^2\Big\lbrack A_+ tan^{-1}\Big(\frac{2}{\gamma}\sqrt{(b+a)/2}\Big)-\frac{A_-}{2}\ln\Big(\frac{\gamma/2+\sqrt{(b-a)/2}}{\gamma/2-\sqrt{(b-a)/2}}\Big)\Big\rbrack \nonumber \\
\end{eqnarray}
with $A_{\pm}=\frac{\gamma^2/4\pm(b\pm a)/2}{\sqrt{(b\pm a)/2}}$, $a=(\omega_c^{na}/2)^2+\omega_0^2-\gamma^2/4$, $b=\sqrt{a^2+(\gamma\omega_c^{na}/2)^2}$ and it is still diamagnetic . For the vanishing $\omega_0\rightarrow 0$ one can obtain
\begin{equation}
M= -\frac{\hbar e}{\pi m c}tan^{-1}\Big(\frac{\omega_c}{\gamma}\Big)
\end{equation}
Now, we can move to the Drude model. With the Drude cut-off the magnetic moment is given by
\begin{equation}
M=-2k_BT\frac{e\omega_c^{na}}{mc} \sum_{n=1}^{\infty}\frac{\nu_n^2(\nu_n+\omega_D)^2}{\lbrack(\nu_n^2+\omega_0^2)(\nu_n+\omega_D)+\nu_n\gamma\omega_D \rbrack^2+\lbrace\omega_c^{na}\nu_n(\nu_n+\omega_D)\rbrace^2}
\end{equation}
 Now, we move to the discussion of momentum-momentum coupling scheme. For the momentum-momentum coupling the magnetic moment is given by
\begin{equation}
M=-2k_BT\frac{e\omega_{cr}^{na}}{mc} \sum_{n=1}^{\infty}\frac{\nu_n^2}{\hat{\lambda}^2(\nu_n)+(\nu_n\omega_{cr}^{na}\hat{G}(\nu_n))^2},
\end{equation}
with $\hat{G}(\nu_n)=G(i\nu_n)=-1+\frac{e}{c}\tilde{u}_{od}(i\nu_n)$ and $\omega_{cr}^{na}=2\phi_0\frac{e}{m_rc}\sqrt{\Lambda_{x,y}^2+\Lambda_{y,y}^2+\Lambda_{z,y}^2}$. So, again we observe diamagnetic moment and it decreases with the increase of non-Abelian parameter $\omega_{cr}^{na}$ and $G(i\nu_n)$.  One can also show that the magnetic moment with Drude cut-off is given by
\begin{equation}
M=-2k_BT\frac{e\omega_{cr}^{na}}{mc} \sum_{n=1}^{\infty}\frac{\nu_n^2(\nu_n+\omega_D)^2}{\lbrack(\nu_n^2+\omega_0^2)(\nu_n+\omega_D)+\nu_n\gamma\omega_D \rbrack^2+\lbrace\omega_{cr}^{na}\hat{G}(\nu_n)\nu_n(\nu_n+\omega_D)\rbrace^2}
\end{equation}
One can observe that the magnetic moment is unaltered in the presence of an arbitrary heat bath. The magnetic moment of the cold atom is still diamagnetic. Also, we find that the free energy increases with the increase of the non-Abelian magnetic field strength.
\section{Conclusions}
In this work, we discuss different quantum thermodynamic properties of a system consisting of a harmonically confined cold atom  in the presence of an artificial uniform vector potential and in contact with a heat bath. We first find the generalized susceptibilities in the coordinate-coordinate coupling scheme. Then we obtain a closed form expression of the free energy of the dissipative cold atom in the presence of underlying background constant vector potential. To demonstrate the effect of non-Abelian term on different QTFs explicitly, we consider the strict Ohmic case and derive the closed form expressions of internal energy U, entropy S and specific heat C. Considering the low temperature expansions of these QTFs, we plot them at low temperatures for three different values of non-Abelian magnetic field strength (N) which clearly show the non-Abelian effect. In this context we also consider the more realistic Drude model heat bath and derive closed form expressions of different QTFs.\\
\indent
Then we derive the generalized susceptibilities in the momentum-momentum coupling scheme. Then, considering arbitrary heat bath spectrum, we derive low temperature expressions of different QTFs. We explicitly show the effect of non-Abelian terms on these QTFs by plotting them for different non-Abelian magnetic field. \\
Finally, we obtain the equilibrium free energy and the magnetic moment of the cold atom due to its internal degrees of freedom. The non-Abelian effect on the magnetic moment for different heat bath is explicitly demonstrated.
\begin{ack}
MB acknowledge the financial support of IIT Bhubaneswar through seed money project SP0045. AMJ thanks DST, India  for award of J C Bose national fellowship.
\end{ack}
{
\end{document}